# The Tetrel Bond

# Definition of the Tetrel Bond

**Pradeep R. Varadwaj[1,2],*  Arpita Varadwaj [1],* Helder M. Marques [2] and Koichi Yamashita[1]**

[1]   Department of Chemical System Engineering, School of Engineering, The University of Tokyo 7-3-1, Tokyo 113-8656, Japan
[2]   Molecular Sciences Institute, School of Chemistry, University of the Witwatersrand, Johannesburg 2050, South Africa
*     Correspondence: pradeep@t.okayama-u.ac.jp (PRV); varadwaj.arpita@gmail.com (AV)

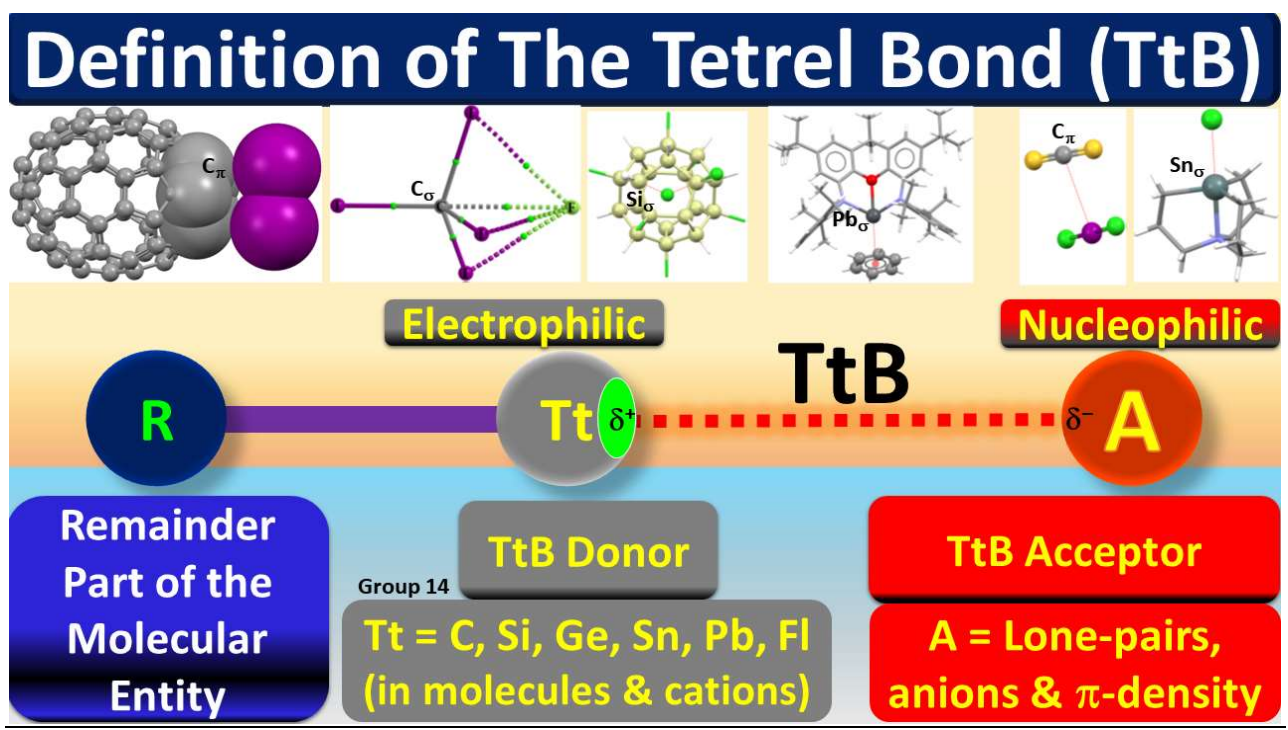

This paper proposes a definition of the term "tetrel bond" based on the evidence documented in the current literature. It lists the donors, acceptors, as well as related characteristic features of tetrel bonds that are commonly observed in the crystalline phase and/or that emerge-from first principles calculations in the solid-state and gas phases. These may be used to identify, characterize and classify the unique subset of inter- and intramolecular interactions formed by the elements of Group 14 of the Periodic Table that possess an electrophilic site in a molecular entity.

**Introduction**

The term "tetrel bond", TtB, has been widely used in the chemistry, crystal engineering, biology, materials science, catalysis, photovoltaics, medicinal and supramolecular chemistry, drug discovery and biomolecular design literature. It is used to describe the attractive non-covalent interactions that are formed when Group 14 elements in molecular entities act as electrophilic sites for nucleophilic sites in the solid, liquid, and gas phases, leading to the formation of complex chemical systems. A significant volume of research has been undertaken focusing on the nature of tetrel bond donors (TtBD) and acceptors (TtBA) that engage attractively to form tetrel bonds.

In this paper, we propose a definition of the term "tetrel bond", together with a list of 15 Notes that clarify the physical and chemical essence of the bond. We list relevant experimentally and theoretically observed features (*a–q*) that appear extensively in various research papers, overviews and reviews. We also enumerate a fairly comprehensive list of possible donors and acceptors of the tetrel bond. In conjunction, we give examples of chemical systems where a tetrel bond may or may not be readily understood – a novel aspect that was not visually shown in any of the three IUPAC proposals that have appeared for hydrogen bonding,[1] halogen bonding[2] and chalcogen bonding.[3] We suggest that the definition be used to specify the subset of inter- and intramolecular non-covalent interactions formed by the elements of the tetrel (Tt) family, which are the elements of Group 14 of the Periodic Table.

The paper complies with the recommendations of IUPAC for the hydrogen bond (HB),[1] the halogen bond (XB),[2] the chalcogen bond (ChB,[3] and our recently-proposed definition of the pnictogen bond (PnB).[4] The definition of the hydrogen bond recommended by IUPAC was revised in 2011; the definition incorporates a variety of features, footnotes and properties.[1] A very similar approach was adopted for XB[2] and ChB,[3] and a somewhat modified approach was used for the proposed definition of the PnB.[4] Our proposed definition of the TtB follows the same pattern of unifying terms for chemical bonds, as was done for the PnB,[4] but with a limited number of features.

Carbon is the lightest element of the Tt family. It has the highest electronegativity and lowest polarizability of the members of this family.[5] It adopts a tetrahedral, planar or linear geometry depending on whether it is bonded to four, three or two other atoms, respectively. In the widely used language of hybridization theory,[6,7] the valence orbitals of carbon are, respectively, $sp^3$, $sp^2$ or $sp$ hybridized. Well-known examples include $CH_4$, $C_2H_4$, HCCH (or HCN), respectively. As will be shown in this paper, a covalently bonded carbon without and with π bonds in molecular entities can act as a nucleophile and is capable of acting as an acceptor of, *inter alia*, hydrogen bonds, halogen bonds, chalcogen bonds and pnictogen bonds. The heavier members of the tetrel family in molecular entities can also act as nucleophiles, depending on their chemical environment; an example is double-bonded Si in molecular entities such as a phosphine-stabilized germasilenylidene, featuring a Ge=S: unit [8] and an NHC-stabilized bromo(silyl)silylene, (R)(Br)$_2$Si=Si(Br)(R').[9]

Methyl groups, when bonded to atoms that are less electronegative than carbon, can act as a Lewis base for the formation of non-covalent interactions;[10, 11] however, when bonded to an electron-withdrawing entity such as F, CN, Cl, or NO$_2$, the carbon atom behaves as an electrophile (as in CF$_4$).[12] Tt = Si, Ge, Sn or Pb (and presumably Fl) in TtH$_4$, as well as in their halogenated derivatives, is electrophilic.[13] In such cases, Tt may be capable of attracting an electron-density rich (nucleophilic) site in the same or separate molecular entity when in close proximity.

In delocalized arene systems, carbon can be either electrophilic or nucleophilic, depending on what it is bonded to, and the chemical environment involved. Specifically, in a simple benzene or benzenol molecule,[14] the centroid region of the delocalized ring formed by the six carbon atoms is nucleophilic. If this benzene ring is bonded to electron-withdrawing substituents such as F, CN, NO$_2$ or C$_6$F$_5$, the centroid region becomes electrophilic.[14-16] There are many compounds known in which carbon exhibits both electrophilic and nucleophilic properties. Allotropic carbons, such as fullerene derivatives, two-dimensional graphene and its nanoflakes, and single- and multi-walled carbon nanotubes, feature both nucleophilic and electrophilic sites.

The benzene analogues (i.e., Hückel structures with $4n + 2$ cyclically delocalized $p$-electrons) of the higher members of the tetrel family are not known crystallographically, although there have been theoretical predictions of aromatic forms of Si such as hexasilabenzene, for example;[17] it is predicted to be non-planar and its chemical properties have yet to be fully elucidated. Aromatic species containing a heavier Group 14 element are known; examples include silabenzene, germabenzene, 1- and 2-silanapthalenes, 9-silaanthracene and phenylsilabenzene.[18]

Whether a covalently or coordinately bonded Tt atom in a molecular entity acts as a nucleophilic or electrophilic site depends on its electronic structure profile. When it acts as an electrophile, the Tt atom can be oriented attractively toward a neighboring nucleophilic site, thus forming a linear or quasi-linear non-covalent interaction. If the entire electrostatic surface of the bound Tt atom in a molecular entity is electrophilic, this may lead to the formation of a non-linear or bent attractive interaction with a nucleophilic site. The term "tetrel bond" refers to the type of non-covalent interaction in which the tetrel atom acts as the electrophile. The presence of electrophilic and nucleophilic sites on the tetrel atom in molecular entities can be identified by experimental and theoretical methods.

We begin by defining the term "tetrel bond", which is given below. This is followed by a number of explanatory Notes that should assist in identifying a TtB. Also provided are a number of illustrative (rather than exhaustive) examples of TtB donors and acceptors.

A review of the literature of experimental and theoretical features of TtBs led to the formulation of this proposal. While we appreciate that the review is by no means comprehensive, we believe that this definition and its attendant features will assist in identifying, characterizing and classifying TtBs in a variety of chemical systems.

**Definition of the Tetrel Bond**

A tetrel bond occurs in chemical systems *when there is evidence of a net attractive interaction between an electrophilic region associated with a covalently or coordinately bonded tetrel atom (or atoms) in a molecular entity and a nucleophilic region in another, or the same, molecular entity.*

**Recommended Notes for Recognition of the Tetrel Bond**

Note 1: A tetrel bond is a non-covalent interaction, represented geometrically by the motif R–Tt⋯A, where Tt is the TtB donor, a tetrel atom (possibly hypervalent) with an electrophilic region on its surface; R is the remainder of the molecular entity R–Tt containing the TtB donor; the TtB donor entity is either electrically neutral, or a molecular cation, that may or may not possess a π-system; A is a TtB acceptor, which may be a lone-pair region in a molecular entity, a molecular anion, or an anion, and must carry a nucleophile.

Note 2: A tetrel bond is either intermolecular or intramolecular; it cannot simultaneously possess both characteristics. However, the same TtB donor may simultaneously engage in the formation of both intermolecular and intramolecular interactions depending on the nature of the interacting partner sites involved.

Note 3: At an equilibrium molecular or complex configuration, TtB donors act as electron density acceptors, and TtB acceptors act as electron density donors.

Note 4: The tetrel bond should be viewed as an attractive interaction between TtB donor site Tt and TtB acceptor site A of opposite charge polarity ($Tt^{\delta+}$ and $A^{\delta-}$), a coulombic attraction between them; $\delta^+$ and $\delta^-$ symbolically refer to the local charge polarity on the interacting regions on the electrostatic surfaces of Tt and A, respectively.

Note 5: The tetrel bond should follow a Type-II topology of non-covalent bonding interaction; a Type-II topology, R–Tt⋯A, is often linear or quasi-linear (but may be non-linear) and satisfies the charge polarity condition (Note 4).

Note 6: The electrophilic region on the TtB donor site Tt is usually observed along the extension of the R–Tt covalent or coordination bond; its exact location on Tt may vary depending on the nature of R. It may appear linearly along or off the outer extension of the R–Tt bond axis; it is conceivable that the stronger the electron-withdrawing capacity of R the greater the electrophilicity of TtB donor Tt.

Note 7: In a tetrel bond the electrophilic region of the TtB donor Tt, generally the region with the lowest (largest deficiency of) electron density, interacts with a region of high electron density on the TtB acceptor A; Tt and A exhibit different directional features and complementarity depending on their surrounding environments.

Note 8: A tetrel bond may occur between the TtB donor site Tt and the TtB acceptor site A within a neutral molecule or between two neutral molecules in close proximity; or it can occur between a neutral molecule with a TtB donor site Tt and an anion (or negative π-density) containing A; or between the TtB donor site Tt in molecular cation and a nucleophile (or negative π-density) A on a neutral molecule; or between an electron-poor delocalized region (positive π-density) as the TtB donor site and a nucleophile A (or delocalized negative π-density) on the TtB acceptor entity containing A; or between two molecular entities of opposite charge polarity with a TtB donor site Tt in the cation and a TtB acceptor in the anion (i.e., an ion-pair).

Note 9: Because of its hypervalent character, a tetrel atom in a molecular entity may concurrently form multiple tetrel bonds with more than one TtB acceptor sites A in the participating molecular entities.

Note 10: The variable electrophilic character of a tetrel atom in a molecular entity may allow it to engage in a number of interactions that lead to the appearance of a variety of electronic and geometric features. The term tetrel bond should not be used to represent attractive interactions where the tetrel atom (frequently carbon) functions as a nucleophile.

Note 11: The electrophilicity, nucleophilicity, and TtB-forming ability of covalently or coordinately bound tetrel atoms may be captured from the sign of the local minima and maxima of the potential on the electrostatic surface of Tt in a molecular entity. The electrophilic region on the electrostatic surface of the covalently or coordinately bound tetrel atom of an isolated molecular entity is often (but not necessarily) represented by a positive (local maximum of) potential. The positive sign of the local maximum of potential may be used as a signature to search for a σ-type tetrel bonds between the TtB donor site Tt and nucleophilic regions on the TtB acceptor site A.

Note 12: If the polarity of the local most maximum or minimum of electrostatic potential is positive at the center of a delocalized double/triple bond or a delocalized ring formed by three or more tetrel atoms, it can behave as a TtB donor site to form a π-type tetrel bond when in proximity to a TtB acceptor site A. Such electrophilic sites are often found around the centroid region of aromatic systems such as fullerenes, carbon nanotubes, and fully halogenated benzene derivatives.

Note 13: Two tetrel sites in two different molecular entities in close proximity may be involved in an attractive engagement to form a tetrel bond; in this case, one of the tetrel atoms acts as a TtB donor site and the other as a TtB acceptor site. Examples include $H_3FTt\cdots\pi(C\equiv C)H_2$ (Tt = Sn, Pb), $H_3XTt\cdots\pi(C=C)H_4$ (Tt = C, Sn, Pb), $H_3XTt\cdots\pi[C_5H_5]^-$ (Tt = Sn, Pb)[19] and $H_3XTt\cdots CMH_3$ (Tt = Si, Ge, Sn, Pb; M = Li, Na, K),[20] and there are many others.[21]

Note 14: The σ- or π-hole on the covalently bonded Tt moiety in molecular entities can be electrophilic or nucleophilic. However, when a σ-hole (or a π-hole) behaves as an electrophile and engages attractively with nucleophiles on the interacting partner molecular entity(s), the resulting interaction between them is either a σ-hole (or a π-hole)-centered non-covalent interaction. Thus, a TtB can be either a σ- or a π-hole interaction depending on whether the TtB donor interacting with the nucleophile features a σ-hole or a π-hole on it.

Note 15: The carbon atom in molecular entities can be electrophilic which can attract the nucleophile on the same or another molecular entity; hence the name "carbon bond" has appeared in the literature. The same nomenclature, in principle, is applicable to the other members of the tetrel family when they possess an electrophilic site and have the capacity to form an intermolecular molecular interaction, as found for the members of the pnictogen family.[11, 22-26] However, for the sake of terminological uniformity, it is recommended that the term "tetrel bond" be used regardless of the nature of involvement of the elements of the tetrel family.

**The Characteristic Features of the Tetrel Bond**

There is experimental evidence (*inter alia* from X-ray diffraction studies, and spectroscopic observations using infrared, Raman, and NMR) of the presence of tetrel bonds in chemical systems. Moreover, signatures that arise from a variety of theoretical studies are also extremely useful. The following is a non-exhaustive list of some distinguishing features that are useful indicators of the occurrence of TtBs in chemical systems. Clearly, the more of these features that are satisfied, the more confident one can be in characterizing a non-covalent interaction as a TtB.

On the formation of a typical tetrel bond R–Tt⋯A between two interacting entities R–Tt and A:

*a.* a coulombic attraction occurs between the interacting regions on the TtBD Tt and TtBA A moieties;

*b.* the sign of the binding energy (or the interaction energy, complexation energy, or stabilization energy) is likely to be negative to signify an energetically favorable interaction;[18-20, 27]

*c.* an energy decomposition analysis, for instance, using Symmetry Adapted Perturbation Theory[28, 29] may indicate the energetic contributions to the binding energy arise from electrostatics, exchange repulsion, polarization (induction and/or charge transfer), and long-range dispersion, and are collectively responsible for the net stabilization of the TtB;

*d.* the separation distance between the TtBD site Tt and the TtBA site A tends to be smaller than the sum of the van der Waals radii of the respective interacting atomic basins but larger than the sum of their covalent bond radii; deviation from the former condition is likely since known van der Waals radii of atoms are only accurate to within ±0.2 Å;[11, 26, 30, 31]

e. the TtB donor tends to approach the TtB acceptor along the outer extension of a σ covalent or coordinate bond, R–Tt; the angular deviation from the R–Tt bond extension is often more pronounced in TtBs than in XBs,[2] ChBs[3] and comparable to that of PnBs;[4, 11]

*f.* the angle of interaction, ∠R–Tt⋯A, tends to be linear or quasi-linear when the approach of the electrophilic site on Tt is along the R–Tt σ covalent or coordinate bond extension, but this can be potentially non-linear or bent when the TtB occurs between a π-type orbital of the bonded Tt atom and the nucleophilic region on TtBA,[32] and when secondary interactions are involved;

*g.* when the nucleophilic region on the TtBD A, is a lone pair orbital, or a (negative) π region, the TtBD tends to approach TtBA along the axis of the lone pair, or orthogonal to the π bond plane;[32]

*h.* the distance of the R–Tt covalent/coordinate bond opposite to the TtB (called tetrel bond donor distance) is typically longer than that in the isolated (unbound) TtBD;[33, 34] *i.* the infrared absorption and Raman scattering observables of both R–Tt and A are affected by TtB formation; the vibrational frequency of the R–Tt bond may be red-shifted (or blue-shifted)[34, 35] depending on the extent of the interactions involved, compared to the frequency of the same bond in the isolated molecular entity; new vibrational modes associated with the formation of the Tt⋯A intermolecular pnictogen bond should also be characteristically observed, as for HBs, XBs and CBs;

*j.* the NMR chemical shifts of nuclei in both R–Tt and TtBA are typically affected (with the former increased or decreased[33]), as found for R–X⋯A XBs and R–Ch⋯A ChBs;[19, 36-39] the isotropic coupling constant of the TtB donor tends to decrease relative to that of the isolated TtB donor,[40, 41] as observed for HBs and XBs;[42]

*k.* the UV–vis absorption bands of the TtBD chromophore may experience a shift to longer wavelengths;

*l.* a bond path and a bond critical point between TtBA donor atom Tt and TtBA site A may be found when an electron density topology analysis[27] based on the quantum theory of atoms in molecules (QTAIM)[43] theory is carried out, together with the appearance of other charge density-based signatures;[44-48]

*m.* isosurface volumes (may be colored greenish, blue, or mixed blue-green between TtBA donor atom Tt and TtBA site A, representative of attractive interactions[23-26, 49, 50]), may be seen if a non-covalent index analysis based on reduced density gradients[51-53] is performed; similar isosurface features[27] may also emerge when an independent gradient model (IGM)[52, 53] or an IGM based on Hirshfeld partition of molecular density (IGMH),[54] or an Interaction Region Indicator (IRI)[55, 56], analysis is performed;

*n.* at least some transfer of charge density from the frontier TtBA orbital to the frontier TtBD orbital may occur;[34, 57, 58] when the transfer of charge density between them is significant, the formation of a dative interaction is likely;[27, 34] the occurrence of the phenomena IUPAC recommended for HBs (see Criteria E1 and Characteristic C5 of Ref. [1]) is also applicable to XBs,[59-62] ChBs,[63-65] and PnBs;[4, 66, 67] the nature of charge transfer may be assessable upon computing the second-order perturbation theory based stabilization energy using Natural Bond Orbital analysis;

*o.* the tetrel bond strength typically increases with a given TtBA A, as the electronegativity of Tt decreases in the order C > Si > Ge > Sn > Pb and the electron withdrawing ability of R increases;[13, 19, 39, 68] there may be an exception when the chemical environment alters the nature of the reactivity between the TtB donor and acceptor moieties.[27]

*p.* the tetrel bond strength may increase for a specific TtBA A and remainder R of the R–Tt entity as the polarizability of the tetrel atom increases (Pb > Sn > Ge > Si > C).[20] This is analogous to the effect observed in the case of XB (I > Br > Cl > F),[69, 70] ChB (Te > Se > S > O),[71] and PnB (Bi > Sb > As > P > N);[67] if a secondary interaction (e.g., hydrogen bond, halogen bond, chalcogen bond, tetrel bond, pnictogen bond, etc.) is involved either with the tetrel bond donor or its acceptor, the order of interaction strength may alter;

*q.* the strength of a tetrel bond can be ultra-strong (>> –40.0 kcal mol$^{-1}$),[27, 34, 72] very strong (–25.0 < energy ≤ –40.0 kcal mol$^{-1}$),[27, 34, 72] strong (–25.0 < energy < –15.0 kcal mol$^{-1}$),[19, 27] moderately strong (–5.0 < energy < –15.0 kcal mol$^{-1}$),[27, 73] weak (–3.0 < energy < –5.0 kcal mol$^{-1}$),[13, 27, 73] very weak (–1.0 < energy < –3.0 kcal mol$^{-1}$),[19, 72] or of the van der Waals type (≤ –1.0 kcal mol$^{-1}$).[74]

**Some Experimentally and Theoretically Known Tetrel Bond Donors and Acceptors**

Some TtB donors and acceptors are listed below. The list is not comprehensive but illustrative. Some of these are from theoretical studies reported by a number of research groups. Others are taken from structures deposited in the Cambridge Structural Database (CSD)[75, 76] or the Inorganic Crystal Structure Database (ICSD).[77, 78]

The TtB donor Tt can be:

– the tetrel atom in Tt H$_3$XTt (Tt = C, Si, Ge, Sn, Pb; X = halogen), TtX$_4$,[27, 79] TtH$_3$NH$_2$[41] (Tt = C, Si, Ge, Sn, Pb),[80] TtH$_4$ (Tt = Si, Ge, Sn, Pb), and in TtX$_2$ (Tt = Sn, Pb),[81] TtF$_3$X (Tt = C, Si, Ge, Sn; X = Cl, Br, I)[81] and TtH$_3$OH (Tt = C, Si, Ge; Sn, Pb);[82]

– the carbon in carboranes (CH$_{11}$B$_{11}$),[83] C$_2$F$_4$, FCCF,[12] XCN (X = F, Cl, Br, I),[12, 84] and FCCCN,[11] CO$_2$,[85] and CS$_2$, C(CN)$_4$; C(CN)$_3$X (X = F, Cl, Br, I);[86]

– a positive π system (the centroids of C$_5$ and C$_6$ of aromatic compounds reinforced by more than one highly electron-withdrawing groups (e.g., C$_4$F$_4$N$_2$, C$_6$X$_6$ (X = F, Cl, Br),[15] C$_6$F$_5$Y, C$_6$H$_5$Y (Y = CN, NO, NO$_2$),[14] C$_6$H$_3$(CN)$_3$, pyrene-F$_{10}$, anthracene-F$_9$, and corenene-F$_{12}$, fullerene derivatives (C$_{60}$, C$_{70}$ and C$_{80}$), single- or multiwall nanotubes, etc.); or the tetrel atom in TtH$_{2-n}$F$_n$=CH$_2$ (*n* = 0-2)[32];

– the Si atom in entities such as dodecachlorohexasilinane (Cl$_{12}$Si$_6$),[87] octachloroundeca-cyclo-icosasilane (H$_{12}$Cl$_8$Si$_{20}$),[88] icosachloro-undecacyclo-icosasilanes (Cl$_{20}$Si$_{20}$), dodecabromohexasilinane (Br$_{12}$Si$_6$),[87] undecabromo(tribromosilyl)hexasilinane (Br$_{14}$Si$_7$),[89] dodecaiodohexasilinane (I$_{12}$Si$_6$),[89] etc.;

- the heavier tetrel atom in SiH$_3$CN,[90] phenyltrifluorosilane,[91] GeH$_3$CN, GeH$_3$(CH$_2$)CN, and GeH$_3$(CH$_2$)$_2$CN, dimethylsilicon dicyanide Si(CH$_3$)(CN)$_2$,[92] dimethylgermanium dicyanide Ge(CH$_3$)(CN)$_2$,[92] dimethyltin dicyanide Sn(CH$_3$)(CN)$_2$;[92]

- Pb in Pb(II) complexes of hydrazine based bis-pyridine Schiff base ligands,[93] and of (iso)nicotinohydrazide-based ligands[94, 95] and pyrimidine-2-carboxylate ligand;[96]

- the tetrel atom in the cations CH$_3$SiH$_2$$^+$,[97] TtH$_3$$^+$ (Tt = C, Si, Ge),[98] C$_6$H$_{15}$O$_2$SSi$^+$,[99] C$_{16}$H$_{34}$ClN$_4$O$_2$Si$^+$,[100] C$_9$H$_{21}$GeN$_3$$^{2+}$,[101] C$_9$H$_{31}$O$_3$Sn$_3$$^+$,[102] C$_{24}$H$_{27}$GeN$_2$O$_3$$^+$,[103] C$_{44}$H$_{28}$IN$_4$Sn$^+$,[104] C$_6$H$_{18}$PSn$^+$,[105] C$_9$H$_{18}$NSn$^+$,[106] C$_6$H$_{19}$OSn$_2$$^+$,[107] H$_6$O$_3$Pb$^{2+}$,[108] C$_{12}$H$_{24}$O$_4$PbSe$^{2+}$,[109] C$_{24}$H$_{20}$N$_4$O$_2$Pb$^{2+}$,[110] and the carbon atom in the methyl group in the derivatives of ammonium (for example, CH$_3$NH$_3$$^+$ and [C$_n$H$_{2n+1}$NH$_3$]$^+$ ($n$ = 2, 3, …18)),[49] as well as in carbocation derivatives (e.g., CH$_3$$^+$, C(CH$_3$)$_3$$^+$, etc).

The TtB acceptor entity A can be:

- a lone-pair on an atom in a molecule. There are almost limitless possibilities, for example, halides in HX, CH$_3$X and CX$_4$ (X = F, Cl, Br, I); N in pyridines or amines, or even in N$_2$; O in H$_2$O, CO, CO$_2$, and in an ether, or a carbonyl group; covalently bonded halogens in molecules; As in AsMe$_3$; a chalcogen in a heterocycle such as a thio-, seleno-, and tellurophene derivatives as well as fused polycyclic derivatives thereof; furoxans, 2,5-thiadiazoles N-oxides, sulfoxide, aryl sulfoxides, and tellurazoles N-oxides; O, S and/or N in derivatives of macrocyclic aza- and/or thia- crown-ethers such as 18-crown-6, 18-azacrown-6, 15-crown-5, N-phenylaza-15-crown-5, and 21-crown-7, cyclam, [2.2.2]cryptand, 1,4,7-triazacyclononane, mono-, di- tri-, and tetrathiadibenzo[24]crown-8-ethers, etc.;

- many anions, such as the halide anions X$^-$; OH$^-$; NO$_3$$^-$; CF$_3$SO$_3$$^-$; BF$_4$**;** tetraphenylborate C$_{24}$H$_{20}$B$^-$; ClO$_4$$^-$; 5-oxotetrazole CHN$_4$O$^-$; I$_3$$^-$; Br$_3$$^-$; N$_3$$^-$; BF$_4$$^-$; AuCl$_4$$^-$; PF$_6$$^-$; AsF$_6$$^-$**;** pentazolide N$_5$$^-$; 5,5′-bistetrazolates C$_2$N$_8$$^{2-}$; $p$-tosylate C$_7$H$_7$SO$_3$$^-$; hexabromo-pentamethyl-1-carbadodecaborane C$_6$H$_{16}$B$_{11}$Br$_6$$^-$; closo-dodecaborate H$_{12}$B$_{12}$$^{2-}$; polyatomic oxyanions such as C$_2$O$_4$$^{2-}$; ReO$_4$$^-$; AsCl$_4$$^-$; SbF$_6$$^-$; SbCl$_4$$^-$; BiCl$_4$$^-$; GaCl$_4$$^-$; ZnCl$_4$$^{2-}$; SiX$_6$$^{2-}$, GeX$_6$$^{2-}$, SnX$_6$$^{2-}$ and PbX$_6$$^{2-}$ (X = F, Cl, Br); PbI$_6$$^{4-}$, SnCl$_3$$^-$, SnI$_3$$^-$, GeCl$_3$$^-$, GeBr$_3$$^-$, GeI$_3$$^-$, Br$_4$Sn$^{2-}$, I$_4$Sn$^{2-}$, Cl$_4$Pb$^{2-}$, I$_5$Pb$^{3-}$, etc.

- a (negative) π system (species featuring a double or triple bond (viz. midpoints of the C=C and C≡C bonds in H$_2$C=CH$_2$ and HC≡CH), and arene moieties of any kind, such as the centroid of the arenes (viz. C$_6$H$_6$, C$_6$H$_5$OH, C$_6$H$_5$F, C$_6$H$_5$NHCH$_3$, C$_6$H$_5$NHOH, and C$_6$(NH$_2$)$_6$), the C=C bonding regions in fullerene C$_{60}$ and C$_{70}$, and N and O in NO$_3$$^-$, etc.).

**Illustrative Crystal Systems *Not* Featuring Tetrel Bonding**

There are many crystallographically-characterized systems where a tetrel atom participates attractively with various elements of the partner molecular entities to provide stability to the resulting crystal lattice. However, the tetrel atom in many of these systems is not electrophilic (but nucleophilic) and is therefore not involved in the formation of tetrel bonds. A few such examples are shown in Figure 1, where carbon participates in forming an attractive intermolecular interaction with an element in a neighboring molecular entity. As such, the attractive intermolecular interaction between H and (C≡C) in the 1,3-diethynylbicyclo[1.1.1]pentane crystal (Figure 1a), as well as of H (–NH$_2$) and (C≡C) in bis(2,4-diamino-6-methyl-1,3,5-triazinium) acetylenedicarboxylate, Figure 1b, is not a tetrel bond; it is a hydrogen bond since the mid-point of (C≡C) is electron-density rich (nucleophilic) that attracts the electrophile on covalently bonded H in the partner molecular entity. The S⋯(C≡C) attractive interaction between sulfur and the midpoint of the –C≡C– bond in bis(phenylethynyl) sulfide, Figure 1c, is not a tetrel (carbon) bond; it is a chalcogen bond since the covalently bonded chalcogen atom has an electrophile that attracts a nucleophile on the –C≡C– triple bond and the intermolecular interaction is directional. The X⋯(C≡C) (X = I, Br) attractive interaction in 1-bromo-2-(iodoethynyl)ferrocene (Figure 1d), 1,4-bis(iodoethynyl)cyclohexane-1,4-diol (Figure 1e), 3-(6-bromo-4-oxo-4*H*-chromen-3-yl)-*N*-(prop-2-yn-1-yl)propenamide (Figure 1f), 1,4-bis(dibromomethylidene)cyclohexane (Figure 1g), and C$_{70}$ Fullerene di-iodine (Figure 1h) is not a carbon-centered tetrel bond; it is a halogen bond since the covalently bonded halogen atom (I or Br) has a positive region (electrophile) that interacts quasi-linearly with the electron-density rich region on the midpoint of the C=C/C≡C double/triple bonds. Notes 1-8 and 11 are not satisfied in order to recognize these interactions as TtBs.

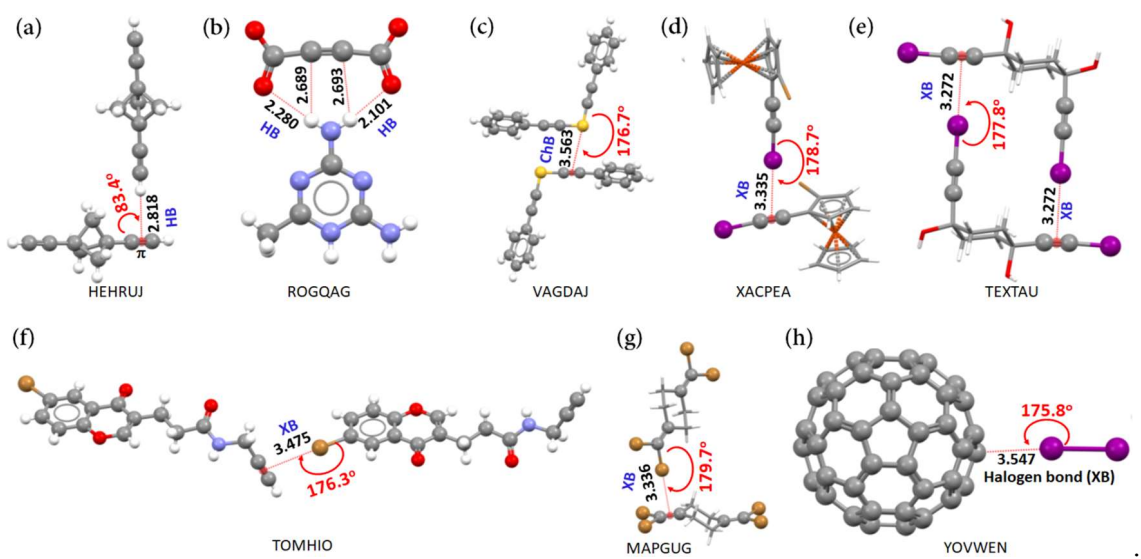

**Figure 1.** Attractive non-covalent interactions between building blocks found in some crystalline materials. (a) 1,3-diethynylbicyclo[1.1.1]pentane;[111] b) bis(2,4-diamino-6-methyl-1,3,5-triazinium) acetylenedicarboxylate;[112] c) bis(phenylethynyl) sulfide;[113] d) 1-bromo-2-(iodoethynyl)ferrocene;[114] e) 1,4-bis(iodoethynyl)cyclohexane-1,4-diol;[115] f) 3-(6-bromo-4-oxo-4*H*-chromen-3-yl)-*N*-(prop-2-yn-1-yl)propenamide;[116] g) 1,4-bis(dibromomethylidene)cyclohexane;[117] h) C$_{70}$ fullerene di-iodine.[118] Selected intermolecular bond angles and bond lengths are given in degrees and Å, respectively. Dotted red lines between interacting atomic entities represent an attractive interaction. Atoms are shown as balls: carbon – gray; nitrogen – blue; hydrogen – white; sulfur – yellow; iron – orange; oxygen – red; bromine – dark yellow; iodine – purple. The CSD reference in uppercase letters is shown in each case.

Many crystals have been reported in which the nature of the non-covalent interaction between interacting moieties cannot be readily captured. Some of them are shown in Fig. 2. Looking at each one, it may be immediately apparent at first glance that there is no tetrel bond in the crystal because the close-contact appears between the negatively charged species (anions). The conclusion may be drawn that the Pb⋯Cl attractive interaction involving lead and chlorine between [Cl$_4$Pb$^{2-}$] building blocks in crystalline [8(C$_9$H$_{14}$N$^+$),Cl$_8$Pb$_2^{4-}$,2$n$(Cl$_4$Pb$^{2-}$)], Figure 2a, is not a tetrel bond; it may be a tetrel-centered lump-hole interaction between sites of unequal charge density. The Ge⋯π(C$_6$) attractive interaction involving the germanium and the centroid of the C$_6$ aromatic ring between a pair of [(C$_{30}$H$_{33}$GeN$_7$)$^{2+}$] dications, Figure 2b, in crystalline [(C$_{30}$H$_{33}$GeN$_7$)$^{2+}$,(CF$_3$O$_3$S)$^{2-}$] may not be a tetrel bond; it too may be a tetrel-centered lump-hole interaction between sites of unequal charge density. Similarly, the Ge⋯Br (Figure 2c), Sn⋯I (Figure 2d), Pb⋯Cl (Figure 2e), Ge⋯I (Figure 2f), Sn⋯I (Figure 2g), and Sn⋯Cl (Figure 2h) close contacts in low-dimensional crystalline halide perovskites [2(C$_8$H$_{11}$FN$^+$),Br$_3$Ge$^-$,Br$^-$], [2(C$_8$H$_{12}$N$^+$),I$_4$Sn$^{2-}$], [2(C$_8$H$_{18}$N$^+$),Cl$_4$Pb$^{2-}$], [C$_6$H$_{11}$N$_2^+$,GeI$_3^-$], [C$_2$H$_7$N$_2^+$,I$_3$Sn$^-$] and [C$_3$H$_{10}$N$^+$,Cl$_3$Sn$^-$], respectively, may not be TtBs; they may represent to tetrel-centered lump-hole interactions between sites of unequal charge density, in which, the tetrel atom donor site is probably relatively less nucleophilic (hole) than the interacting halogen (or π site) (a lump) in the interacting molecular entity. We note that the intermolecular distances corresponding to the close contacts Pb⋯Cl (Figure 2a,e), Ge⋯π(C$_6$) (Figure 2b), Ge⋯Br (Figure 2c), Sn⋯I (Figure 2d, g), and Sn⋯Cl (Figure 2h) are close to, or less than, the sum of the respective van der Waals radii of bonded atomic basins, satisfying characteristic criterion $d$ of TtBs. The ∠X–Tt⋯X (X = halogen) and ∠N–Tt⋯π(C$_6$) angular features satisfy characteristic criteria $e$ and $f$ of TtBs. However, because the occurrence of either of these non-covalent interactions lacks an electrophile on the Tt site on the anion moieties, and Notes 1–13 are not readily satisfied, they may not be regarded as TtBs.

The occurrence of tetrel-centered close contacts between anions in low-dimensional crystalline organic-inorganic halide perovskites is not surprising. They appear in addition to a large number of H⋯X hydrogen bonds and other interactions that interplay to provide structural stability, and hence are collectively responsible for the functionality of organic-inorganic hybrid materials. In other words, hydrogen bonds are arguably the key players in the arrangement and formation of inorganic layers by assembling the anionic moieties (such as Cl$_4$Pb$^{2-}$, Br$_3$Ge$^-$, GeI$_3^-$ and I$_3$Sn$^-$) through tetrel-centered close contacts and hence are responsible for the development of the structure of hybrid organic-inorganic metal halide perovskites.[25] It is also likely that the halogen derivative(s) in the anion (e.g., SnCl$_3^-$) donates some charge to the antibonding orbitals of the interacting amine and methyl H atoms in the process of hydrogen bond formation with the interacting cation(s). This is related to the polarization of the charge density on the surface of the tetrel derivative(s) in the anion, which appears as charge density deficient regions along the X–Tt bond extensions. The positive regions so-developed on the surface of the Tt atoms are adequate to form tetrel bonds with the negative sites on the halogen derivatives of the interacting anion(s).

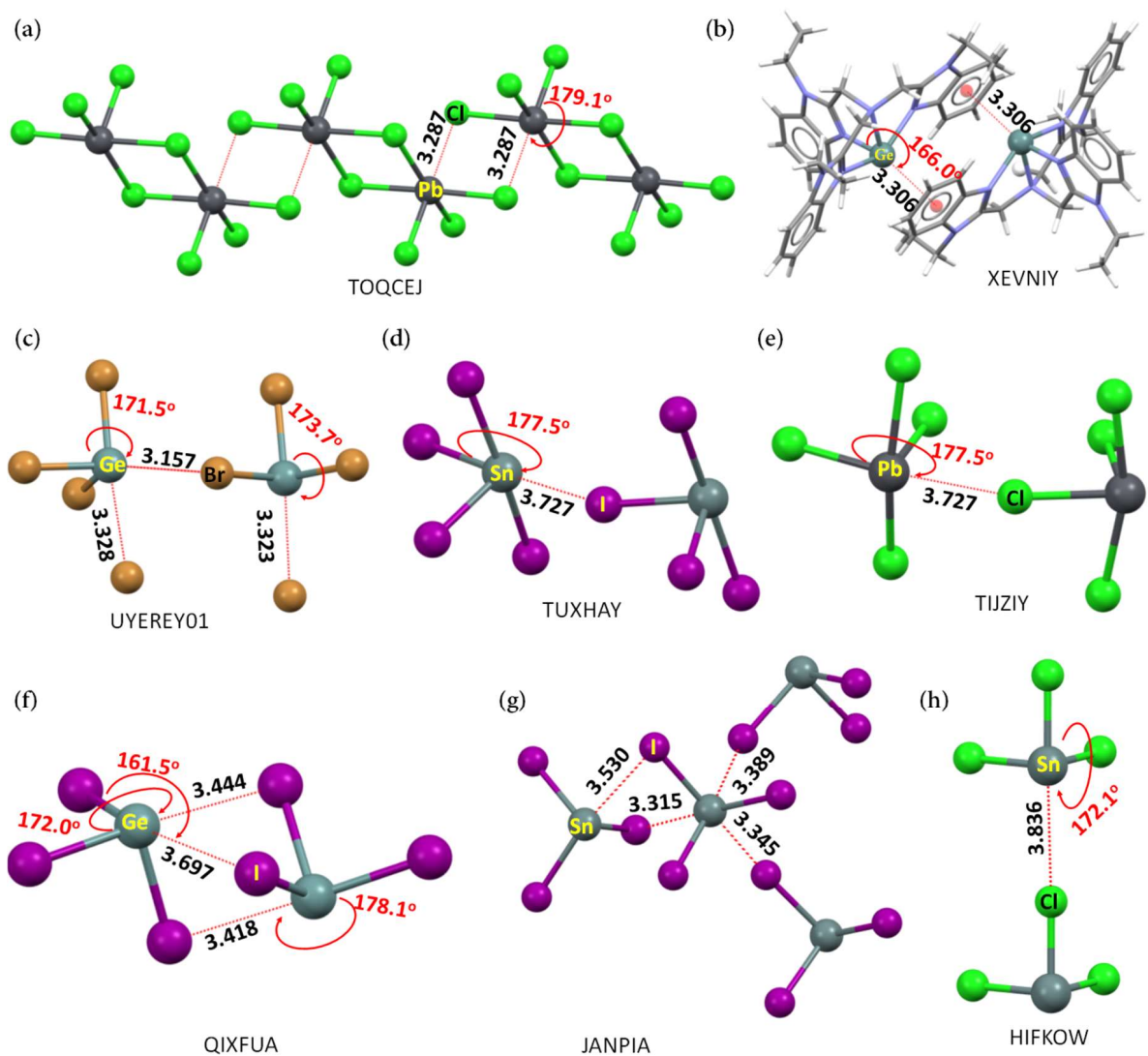

**Figure 2.** Attractive non-covalent interactions between building blocks of similar charge polarity found in some crystalline adducts. (a) catena-[octakis(2,4,6-trimethylanilinium) bis(bis(μ-chloro)-dichloro-lead) bis(μ-chloro)-hexachloro-lead];[119] (b) (*N,N,N*-tris[(1-ethyl-1*H*-benzimidazol-2-yl)methyl]amine)-germanium bis(trifluoromethanesulfonate);[120] (c) bis((2-(4-fluorophenyl)ethyl)ammonium) bromide tribromo-germanium(II);[121] (d) bis((*R*)-1-phenylethan-1-aminium) tetrakis(iodo)-tin;[122] (e) bis((*R*)-1-cyclohexylethylammonium) tetrachloro-lead(II);[123] (f) 1-ethyl-3-methyl-1*H*-imidazol-3-ium tris(iodo)germanate;[124] (g) acetamidinium triiodo-tin(II);[125] (h) trimethylammonium trichloro-tin(IV).[126] Selected intermolecular bond angles and bond lengths are given in degrees and Å, respectively. Dotted lines between interacting atomic entities represent an attractive interaction. Atoms are shown as balls: carbon – gray; nitrogen – blue; hydrogen – white; chlorine – green; germanium – dark green; lead – dark gray. The CSD reference is shown for each case. Building blocks such as 2,4,6-trimethylanilinium in (a), trifluoromethanesulfonate in (b), and the cations in (c)-(h) are omitted for clarity.

**Illustrative Crystal Systems Featuring Tetrel Bonding**

The C···π(C=C) attractive engagement between the methyl carbon in 1-methylcycloprop-2-ene-1-carboxylic acid and the midpoint of the C=C fragment in a neighboring unit, Figure 3a, is a σ–π type tetrel bond between two neutral molecules. The $C_\pi$···O and $C_\pi$···Br attractive contacts between the building blocks of neutral molecular entities in crystalline [$C_3HN_3O_5$], 3,5-dinitroisoxazole, (Figure 3b) and in 2-(bromomethylidene)pentanedioic acid (Figure 3c) are π–lone-pair type tetrel bonds. The $C_\pi$···I attractive interaction between the carbon atom in $CS_2$ and the iodine atom in $ICl_2^-$, Figure 3d, is a π–anion type tetrel bond in crystalline [$C_{64}H_9N_2^+$,$Cl_2I^-$,$CS_2$]. The $C_\pi$···N close contact between the central carbon atom of the fragment –C–C≡N in the [$C_{13}H_{12}N_2O_2$] unit and the nitrogen atom in the same fragment of a neighboring unit in crystalline [$C_{13}H_{12}N_2O_2$], Figure 3e, is a π–lone-pair type tetrel bond. The Si···Cl attractive interactions between the silicon atoms in [$H_{12}Cl_8Si_{20}$] and the encaged $Cl^-$ anion in the crystal of [$2(C_{36}H_{30}NP_2^+)$,$H_{12}Cl_8Si_{20}$,$C_7H_7O_3S^-$,$Cl^-$], Figure 3f, are σ–anion type charge-assisted tetrel bonds. The C···π(C=C) attractive contact between the building blocks ($C_{120}^{2-}$) and ($C_5H_9N_2^+$) of ionic entities in crystalline [$2(C_{10}H_{20}CdIN_2S_4^-)$,$C_{120}^{2-}$,$4(C_5H_9N_2^+)$] (Figure 3g) is a cation-anion σ–π type tetrel bond. The C···π(C=C) attractive interaction between the building blocks in crystalline [$C_{69}H_{22}Pd$,$1.5(CS_2)$], Figure 3h, is a σ–π type tetrel bond. Similarly, the Si(σ)···I(lone-pair) close contact between the interacting molecular units in crystalline $(I_{12}Si_6^{2+})(I^-)_2$, Figure 3i, is an Si-centered tetrel bond; it is not a halogen bond. The σ–π (Figure 3a,g,h), and σ–lone-pair (Figure 3f, i) type tetrel bonds are directional, whereas the π–lone-pair interactions are non-directional. They satisfy characteristic features *d-f*, including Notes 1-12.

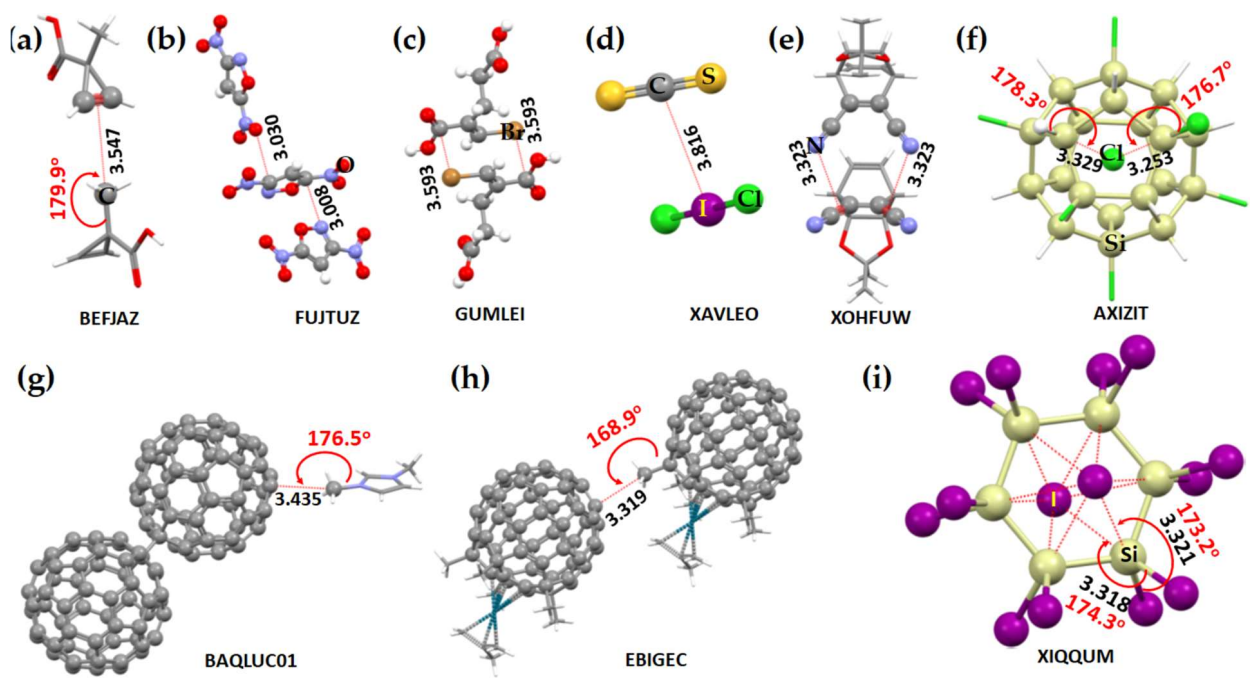

**Figure 3.** Attractive non-covalent tetrel bonds between building blocks found in some crystalline materials. (a) 1-methylcycloprop-2-ene-1-carboxylic acid;[127] (b) 3,5-dinitroisoxazole;[128] (c) 2-(bromomethylidene)pentanedioic acid;[129] (d) (piperazin-1-ium-1,4-diyl)C$_{60}$ fullerene dichloroiodide carbon disulfide (only ICl$_2^-$ and CS$_2$ shown);[130] (e) *syn*-4,4-dimethyl-3,5-dioxatricyclo(5.2.2.0$^{2,6}$)undeca-8,10-diene-8,9-dicarbonitrile;[131] (f) [2(C$_{36}$H$_{30}$NP$_2^+$),H$_{12}$Cl$_8$Si$_{20}$,C$_7$H$_7$O$_3$S$^-$,Cl$^-$];[88] (g) [2(C$_{10}$H$_{20}$CdIN$_2$S$_4^-$),C$_{120}^{2-}$,4(C$_5$H$_9$N$_2^+$)];[132] (h) [C$_{69}$H$_{22}$Pd,1.5(CS$_2$)];[133] (i) bis(tetrabutylphosphanium) dodecaiodohexasilinane bis(iodide) [2(C$_{16}$H$_{36}$P$^+$),I$_{12}$Si$_6$,2(I$^-$)].[89] Selected intermolecular bond angles and bond lengths are given in degrees and Å, respectively. Dotted lines between interacting atomic entities represent an attractive interaction. Atoms are shown as balls: carbon – gray; chlorine – green; nitrogen – blue; sulfur – yellow; iodine – purple; hydrogen – white; oxygen – red. The CSD reference is shown for each case. The (piperazin-1-ium-1,4-diyl)C$_{60}$ fullerene unit in the crystal in (d), the (C$_{36}$H$_{30}$NP$_2^+$) and C$_7$H$_7$O$_3$S$^-$] units in (f), the C$_{10}$H$_{20}$CdIN$_2$S$_4^-$ unit in (g), the CS$_2$ unit in (h), and the (C$_{36}$H$_{30}$NP$_2^+$) unit in (i) have been omitted for clarity. Atom labeling is shown for selected atoms.

Some crystalline fullerene-based systems are shown in Figure 4. The interacting molecular entities in each of these systems are neutral. The (C$_{70}$)π···Cl (Figure 4a), (C$_{70}$)π···S (Figure 4b), (C$_{70}$)π···I (Figure 4c), (C$_{70}$)π···S (Figure 4d), (CS$_2$)C$_\pi$···π(C$_{70}$) (Figure 4e), (C$_{70}$)π···π(benzene) (Figure 4f) attractive intermolecular interactions between participating molecular entities are π-type tetrel bonds. The C···O attractive intramolecular interaction in crystalline methoxy-benzyl-C$_{70}$ fullerene, Figure 4g, is a tetrel bond; however, the C(π)···S attractive intramolecular interaction in crystalline b-(1,2-xylene)-C$_{70}$ fullerene carbon disulfide, Figure 4h, is a not a chalcogen bond, but a tetrel bond.

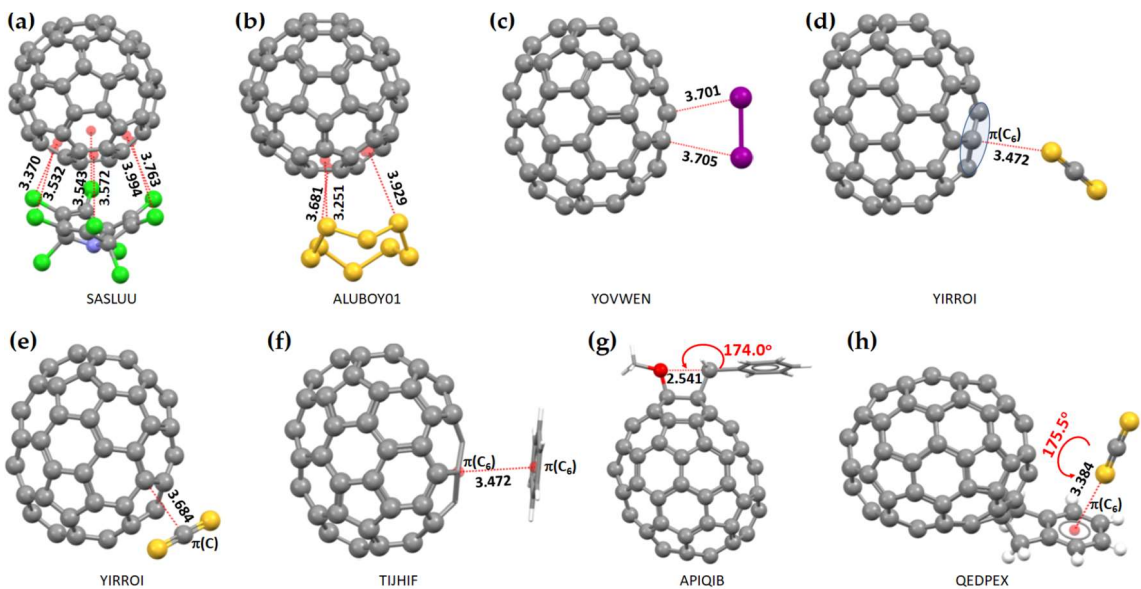

**Figure 4.** Attractive non-covalent interactions between building blocks found in some crystalline fullerene-based materials. (a) nonachloroazatriquinacene C$_{70}$ fullerene;[134] (b) C$_{70}$ fullerene bis(octa-sulfur);[135] (c) C$_{70}$ fullerene di-iodine;[118] (d) C$_{70}$ fullerene carbon disulfide solvate;[136] (e) C$_{70}$ fullerene carbon disulfide solvate;[136] (f) C$_{70}$ fullerene bis(ethylenedithio)tetrathiafulvalene benzene solvate;[137] (g) methoxy-benzyl-C$_{70}$ fullerene;[138] (h) b-(1,2-xylene)-C$_{70}$ fullerene carbon disulfide.[139] Selected intermolecular bond angles and bond lengths are given in degrees and Å, respectively. Dotted red lines between interacting atomic entities represent an attractive interaction. Atoms are shown as balls: carbon – gray; chlorine – green; nitrogen – blue; sulfur – yellow; iodine – purple; hydrogen – white; oxygen – red. The CSD reference in uppercase letters is shown for each case. Bis(ethylenedithio)tetrathiafulvalene in (f) is omitted for clarity.

Several other solid-state structures that feature tetrel bonding are shown in Figure 5. They are charge-assisted, and occur between a pair of molecular ions. The Si···O and Si···Cl close contacts in crystals of [$C_{27}H_{39}N_2Si^+$,$CF_3O_3S^-$] (Figure 5a) and [$C_{24}H_{27}GeN_2O_3^+$,$F_6Sb^-$] (Figure 5b), respectively, are double charge-assisted tetrel bonds; each is formed via attraction between a cation containing electrophilic Si and an anion containing nucleophilic O or Cl. The Ge···F and Ge···O close contacts in crystals of [$C_{24}H_{27}GeN_2O_3^+$,$F_6Sb^-$] (Figure 5c)[103] and [$C_{12}H_{30}As_2Cl_2Ge^{2+}$,$2(CF_3O_3S^-)$] (Figure 5d)[140] are double and quadruple charge-assisted σ–anion type tetrel bonds, respectively, but the Ge···O close contact, Figure 5e, in the crystal of [$C_{12}H_{15}GeN_6P^{2+}$,$C_4H_8O_2$,$2(CF_3O_3S^-)$,$C_2H_3N$][141] is a multiple charge-assisted σ–lone-pair type tetrel bond. The Sn···Cl, Sn···N, Sn···Cl, and Sn···Br/Sn···π(arene) attractive contacts in crystals shown in Figures 5f, 5g, 5h, and 5j are charge assisted tetrel bonds, but each Sn···N attractive interaction in Figure 5i is a triply charge-assisted tetrel bond. The Pb···π(arene) attractive interactions in crystals shown in Figure 5k and 5l, and the Pb···F attractive interactions between ions in [$C_{24}H_{20}N_4O_2Pb^{2+}$] and [$F_6Si^{2-}$]$_2$, Figure 5m, are also charge-assisted tetrel bonds.

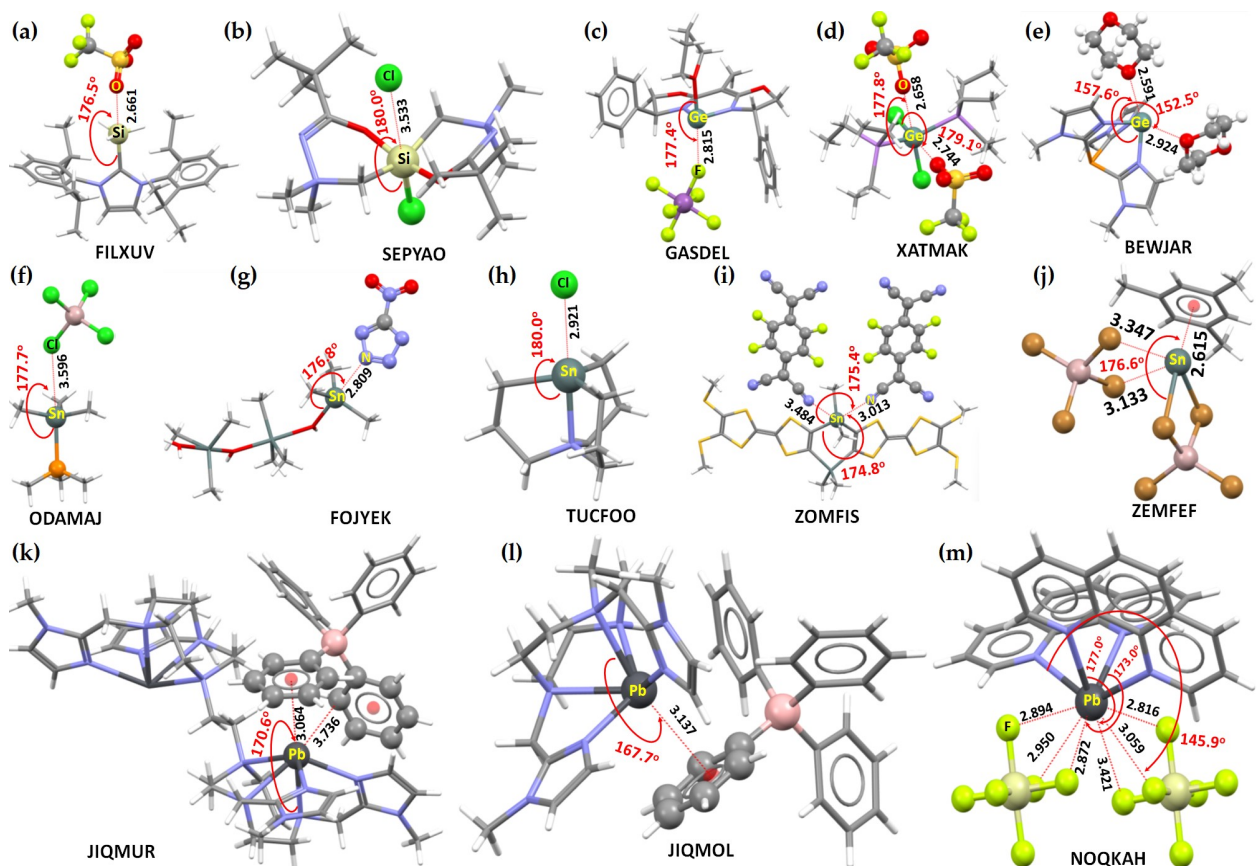

**Figure 5.** Attractive non-covalent TtB interactions between building blocks found in some crystalline materials: (a) (1,3-bis(2,6-diisopropylphenyl)imidazol-2-ylidene)-trihydrido-silylium trifluoromethanesulfonate [$C_{27}H_{39}N_2Si^+$,$CF_3O_3S^-$];[142] (b) bis(*N*,*N*-dimethyl-*N*-(pivaloimidato)ammoniomethyl-C,O)-chloro-siliconium chloride [$C_{16}H_{34}ClN_4O_2Si^+$,$Cl^-$];[100] (c) (4-phenyl-2-(1-(4-phenyl-4,5-dihydro-1,3-oxazol-2-yl)ethylidene)-1,3-oxazolidinato)-(tetrahydrofuran)-germanium(II) hexafluoroantimonate [$C_{24}H_{27}GeN_2O_3^+$,$F_6Sb^-$];[103] (d) dichloro-bis(triethylarsane)-germanium(IV) bis(trifluoromethanesulfonate) [$C_{12}H_{30}As_2Cl_2Ge^{2+}$,$2(CF_3O_3S^-)$];[140] (e) (2,2',2''-phosphanetriyltris(1-methyl-1*H*-imidazole))-germanium(II) bis(trifluoromethanesulfonate) 1,4-dioxane acetonitrile solvate [$C_{12}H_{15}GeN_6P^{2+}$,$C_4H_8O_2$,$2(CF_3O_3S^-)$,$C_2H_3N$];[141] (f) trimethyl-(trimethylphosphine)-tin tetrachloro-aluminium [$C_6H_{18}PSn^+$,$AlCl_4^-$];[105] (g) bis($\mu_2$-hydroxo)-aqua-nonamethyl-tri-tin 5-nitrotetrazolate [$C_9H_{31}O_3Sn_3^+$,$CN_5O_2^-$];[102] (h) tetrakis((tris(propane-1,3-diyl)amine)-tin) tris(hexafluoro-antimony) chloride [$4(C_9H_{18}NSn^+)$,$3(F_6Sb^-)$,$Cl^-$];[106] (i) bis($\mu_2$-2-(4,5-bis(methylsulfanyl)-1,3-dithiol-2-ylidene)-1,3-dithiole-4,5-diyl)-tetramethyl-di-tin radical dication bis(2,3,5,6-tetrafluoro-7,7,8,8-tetracyanoquinodimethane) benzene solvate [$C_{20}H_{24}S_{12}Sn_2^{2+}$,$2(C_{12}F_4N_4^-)$,$C_6H_6$];[143] (j) bis($\mu$-bromo)-dibromo-aluminium-tin tris($\mu$-bromo)-pentabromo-di-aluminium-tin mesitylene solvate [$2(C_9H_{12})$,$Al_2Br_8Sn$,$AlBr_4Sn^+$,$AlBr_4^-$];[144] (k) ($\mu_2$-1,2-bis(4,7-bis(1-methylimidazol-2-yl-methyl)-1,4,7-triazacyclonon-1-yl)ethane)-di-lead(II) tetrakis(tetraphenylborate) acetonitrile solvate [$C_{34}H_{56}N_{14}Pb_2^{4+}$,$4(C_{24}H_{20}B^-)$,$4(C_2H_3N)$];[145] (l) 1,4-bis(1-methylimidazol-2-yl-methyl)-1,4,7-triazacyclononane-lead(II) bis(tetraphenylborate) [$C_{16}H_{27}N_7Pb^{2+}$,$2(C_{24}H_{20}B^-)$];[145] (m) bis(1,10-phenanthroline)-diaqua-lead bis(bis(1,10-phenanthroline)-lead) tris(hexafluorosilicate) undecahydrate [$C_{24}H_{20}N_4O_2Pb^{2+}$,$2(C_{24}H_{16}N_4Pb^{2+})$,$3(F_6Si^{2-})$,$11(H_2O)$].[110] Selected intermolecular bond angles and bond lengths are given in degrees and Å, respectively. Dotted red lines between interacting atomic entities represent an attractive interaction. The CSD reference in uppercase letters is shown for each case. Some building blocks in most of these crystal systems are omitted for clarity.

**Concluding Remarks**

Tetrel bonding is a versatile non-covalent interaction or supramolecular synthon. It is an attractive force that brings together two or more molecular units, leading to a supramolecular assembly in large-scale systems. It is ubiquitous in many chemical systems, including single molecules, molecular complexes, crystalline solids, supramolecular structures, molecular chemistry and biology, and functional nanomaterials such as metal halide perovskites. Its implications in crystal engineering, anion binding, recognition and transport, catalysis and photovoltaics, among other research fields, are significant. Its key effect is to cause (or assist in) the assembly of molecular entities by electrostatic attraction, even though other forces (for example, polarization and dispersion) contribute to the stability of this interaction.

The list of tetrel bond donors and acceptors, as well as the characteristics of the tetrel bond, is vast. In this article, we have highlighted a few. We have provided several illustrative examples that can assist in recognizing instances where tetrel bonding may occur. We have also pointed out pitfalls where a putative TtB is actually a different non-covalent interaction. The definition and characteristic features listed may be useful to assist students and researchers in a variety of research fields to identify, characterize, and classify this unique subset of non-covalent interaction in a variety of chemical systems.

It should be borne in mind that the IUPAC recommended definitions, notes, and characteristic features of hydrogen, halogen, and chalcogen bonds are biased only toward σ-hole centered non-covalent (Type-II) interactions. They do not explicitly clarify, for example, whether they are also applicable to π-hole-centered halogen and chalcogen bonds. This view is justified because it has been demonstrated that either of these non-covalent bonds appears only when an electrophile is observed on the atom Z on the opposite side of the R–Z covalent bond that engages attractively with an electron density donor atom A to form the aforementioned bond, where Z refers to the hydrogen, or halogen/chalcogen derivative. This general view can, in principle, be applied to non-covalent interactions formed by the elements of the main group and other elements of the Periodic Table as well. The "tetrel bond" formed by elements of Group 14 also follows the same guiding principles as other non-covalent bonds. However, it is worth mentioning that the definition, notes and characteristics that apply to σ-hole-centered tetrel bonds may also be applicable to π-hole-centered tetrel bonds when a π-hole appears on the electrostatic surface of an atomic or molecular entity; this is specifically true when the covalently (or coordinately) bound tetrel atom (or atoms) in molecular entities holds an electrophilic π-hole that constructively attracts nucleophile(s) on tetrel bond acceptor site A. Several chemical systems are illustrated in this study that feature the possible occurrence of π-hole-centered tetrel bonds in crystalline materials.

**Author Contributions:** Conceptualization, project design, and project administration, P.R.V.; formal analysis and investigation, P.R.V. and A.V.; supervision, P.R.V.; writing—original draft, P.R.V. and A.V.; writing—review, editing and validation, P.R.V., H.M.M., A.V., and K.Y. All authors have read and agreed to the published version of the manuscript.

**Funding:** This research received no external funding.

**Institutional Review Board Statement:** Not applicable.

**Informed Consent Statement:** Not applicable.

**Data Availability Statement:** This research did not report any data.

**Acknowledgments:** This work was entirely conducted using various laboratory facilities provided by the University of Tokyo and the University of the Witwatersrand. P.R.V. is currently affiliated with the University of the Witwatersrand (SA) and Nagoya University (Japan). A.V. is currently affiliated with Tokyo University of Science (Japan). K.Y. is currently affiliated with Yokohama City University and Kyoto University, ESICB, Kyoto (Japan). H.M.M. thanks the National Research Foundation, Pretoria, South Africa and the University of the Witwatersrand for funding.

**Conflicts of Interest:** The authors declare no conflict of interest. The funders had absolutely no role in the design of the study; in the collection, analyses, or interpretation of data; in the writing of the manuscript; or in the decision to publish the results.